\documentclass[twocolumn,aps,prl,superscriptaddress,amsmath,amssymb,showpacs]{revtex4-1}

\usepackage{graphicx}
\usepackage{natbib,color}

\begin{document}



\title{Spin anisotropy of the resonance in superconducting FeSe$_{0.5}$Te$_{0.5}$}

\author{P.~Babkevich}
\email[]{peter.babkevich@physics.ox.ac.uk}
\affiliation{Department of Physics, Oxford University, Oxford, OX1 3PU, United Kingdom}
\affiliation{Laboratory for Neutron Scattering, Paul Scherrer Institut, CH-5232 Villigen PSI, Switzerland}
\author{B.~Roessli}
\affiliation{Laboratory for Neutron Scattering, Paul Scherrer Institut, CH-5232 Villigen PSI, Switzerland}
\author{S.~N.~Gvasaliya}
\affiliation{Institute for Solid State Physics, ETH Z{\"u}rich, Z{\"u}rich CH-8093, Switzerland}
\altaffiliation[previous address: ]{Laboratory for Neutron Scattering, Paul Scherrer Institut, CH-5232 Villigen PSI, Switzerland}
\author{L.-P.~Regnault}
\affiliation{SPSMS, UMR-E9001, CEA/UJF-Grenoble 1, MDN,
17 rue des Martyrs, 38054 Grenoble Cedex 9, France}
\author{P.~G.~Freeman}
\affiliation{Institut Laue-Langevin, BP 156, 38042 Grenoble Cedex 9, France}
\author{E.~Pomjakushina}
\affiliation{Laboratory for Developments and Methods, Paul Scherrer Institut, CH-5232 Villigen PSI, Switzerland}
\author{K.~Conder}
\affiliation{Laboratory for Developments and Methods, Paul Scherrer Institut, CH-5232 Villigen PSI, Switzerland}
\author{A.~T.~Boothroyd}
\affiliation{Department of Physics, Oxford University, Oxford, OX1 3PU, United Kingdom}

\date{\today}

\begin{abstract}
We have used polarized-neutron inelastic scattering to resolve the spin fluctuations in superconducting FeSe$_{0.5}$Te$_{0.5}$ into components parallel and perpendicular to the layers. A spin resonance at an energy of 6.5\,meV is observed to develop below $T_{\rm c}$ in both fluctuation components. The resonance peak is anisotropic, with the in-plane component slightly larger than the out-of-plane component. Away from the resonance peak the magnetic fluctuations are isotropic in the energy range studied. The results are consistent with a dominant singlet pairing state with $s^{\pm}$ symmetry, with a possible minority component of different symmetry.
\end{abstract}



\maketitle

The discovery of superconductivity in iron pnictides and chalcogenides with transition temperatures $T_{\rm c}$ up to $55$\,K has prompted comparisons with the high-$T_{\rm c}$ copper-oxide superconductors \cite{johnston-advances-2010,lynn-physicac-2009,lumsden-jpcm-2010}. In common with the cuprates, the phase diagram of the Fe-based superconductors shows a suppression of static magnetic order and the emergence of superconductivity with doping. Also like the cuprates, a spin resonance develops below $T_{\rm c}$ in the magnetic spectrum of the Fe-based superconductors as measured by inelastic neutron scattering \cite{christianson-nature-2008,lumsden-prl-2009,chi-prl-2009,qiu-prl-2009,babkevich-jpcm-2010,argyriou-prb-2010,mook-prl-2010}. The existence of a superconductivity-induced spin resonance peak has been shown to relate to the superconducting pairing state and gap symmetry \cite{korshunov-prb-2008,maier-prb-2008}.

Among the Fe-based superconductors, the iron chalcogenides Fe$_y$Se$_x$Te$_{1-x}$ have the simplest crystal structure (space group $P4/nmm$, room temperature lattice parameters $a = b \approx 3.8$\,\AA, $c \approx 6.1$\,\AA\ \cite{yeh-epl-2008}). This, together with the availability of large single crystals, makes them attractive for fundamental studies.  Antiferromagnetic order characteristic of the parent compound Fe$_{1+y}$Te persists up to $x\sim 0.1$, after which short-range magnetic order and partial superconductivity coexist for concentrations $ x \leq 0.5$ \cite{khasanov-prb-2009}. Bulk superconductivity is reported for $x > 0.4$ with a maximum $T_{\rm c}\approx 15$\,K at $x\approx 0.5$ \cite{yeh-epl-2008}.

Inelastic neutron scattering experiments have shown that the spin fluctuations in Fe$_y$Se$_x$Te$_{1-x}$ extend up to 250\,meV \cite{lumsden-nature-2010}. A spin resonance is observed to develop below $T_{\rm c}$ at an energy of 6.5\,meV, centered on wave vectors of the form ${\bf Q}_0 = (0.5,0.5,l)$ \cite{qiu-prl-2009,babkevich-jpcm-2010,argyriou-prb-2010,mook-prl-2010,
wen-prb-2010}. The resonance peak is quasi-two-dimensional, which means that it varies only weakly with the out-of-plane wavevector component $l$ \cite{qiu-prl-2009}. The position of the resonance peak in momentum space carries information about the symmetry of the superconducting state. For example, for singlet pairing, the BCS coherence factor enhances the neutron response function when the superconducting gap changes sign between the points on the Fermi surface connected by ${\bf Q}_0$~\cite{schrieffer-book-1964}. In iron-based superconductors, the singlet $s^{\pm}$ pairing state \cite{mazin-prl-2008} is consistent with many experimental results including the existence of a spin resonance at ${\bf Q}_0$ \cite{christianson-nature-2008}. However, a spin resonance at ${\bf Q}_0$ is not particular to $s^{\pm}$. It is also predicted, for example, for certain triplet $p$-wave states \cite{maier-prb-2008}.

Until now, inelastic neutron scattering measurements on Fe$_y$Se$_x$Te$_{1-x}$ were performed with an unpolarized neutron beam. However, certain superconducting gap functions can result in anisotropic spin susceptibilities at the resonance energy~\cite{maier-prb-2008, joynt-prb-1988}. In this Rapid Communication, we report the results of polarized-neutron inelastic scattering measurements on FeSe$_{0.5}$Te$_{0.5}$ which determine the anisotropy of the imaginary part of the dynamical susceptibility $\chi''(\bf{Q}, \omega)$. We find that the resonance peak exhibits a spin anisotropy such that the in-plane component $ \chi''_{ab}({\bf Q}_0, \omega)$ is larger by about 20\% than the out-of-plane component $ \chi''_{c}({\bf Q}_0, \omega)$. This is consistent with a dominant singlet superconducting ground state with $s^{\pm}$ symmetry, and contrasts with a recent polarized-neutron scattering study of BaFe$_{1.9}$Ni$_{0.1}$As$_2$ which revealed a highly anisotropic spin resonance peak appearing only in the in-plane response \cite{lipscombe-prb-2010}.
%


The single crystal sample of nominal composition of FeSe$_{0.5}$Te$_{0.5}$ was grown by the modified Bridgman method \cite{sales-prb-2009,bendele-prb-2010}. Analysis of pulverized crystals from the same batch by x-ray powder diffraction revealed a composition Fe$_{1.045}$Se$_{0.406}$Te$_{0.594}$ with traces of Fe$_7$Se$_8$ (5\% volume fraction) and Fe ($\le 1$\%) as impurity phases \cite{bendele-prb-2010}. In a previous study \cite{babkevich-jpcm-2010} we performed magnetometry measurements on a piece of the same crystal and found bulk superconductivity below $T_{\rm c}=14$\,K. The neutron scattering sample was rod-shaped and had a mass of approximately 5\,g. The mosaic spread in the $ab$-plane was found to be $1.5^\circ$ (full-width at half-maximum).

%
The inelastic neutron scattering measurements were carried out on the IN22 triple-axis spectrometer at the Institut Laue-Langevin. The crystal was aligned with the $c$ axis perpendicular to the scattering plane and mounted in an ILL-type orange cryostat. The spectrometer was operated with a fixed final wavevector of $k_{\rm f}= 2.66$\,\AA$^{-1}$ and without collimation. A graphite filter was installed in the scattered beam to suppress contamination by higher-order wavelengths. The analyzer was horizontally-focused to increase intensity. The corresponding energy resolution with this setup is approximately 0.8\,meV at the elastic position.
%
Longitudinal polarisation analysis was performed with the Cryopad device \cite{regnault-physicab-2003}.  Cryopad is designed such that the sample is in a zero magnetic-field environment, and the incident and final neutron polarization states are controlled with nutation and precession fields which are decoupled by superconducting Nb shielding.  With a Heusler monochromator and analyzer the effective flipping ratio was about 10 as measured on the $(110)$ structural Bragg peak. No corrections were made to compensate for the non-ideal polarization.

\begin{figure}
\centering
\includegraphics[width=0.8\columnwidth]
{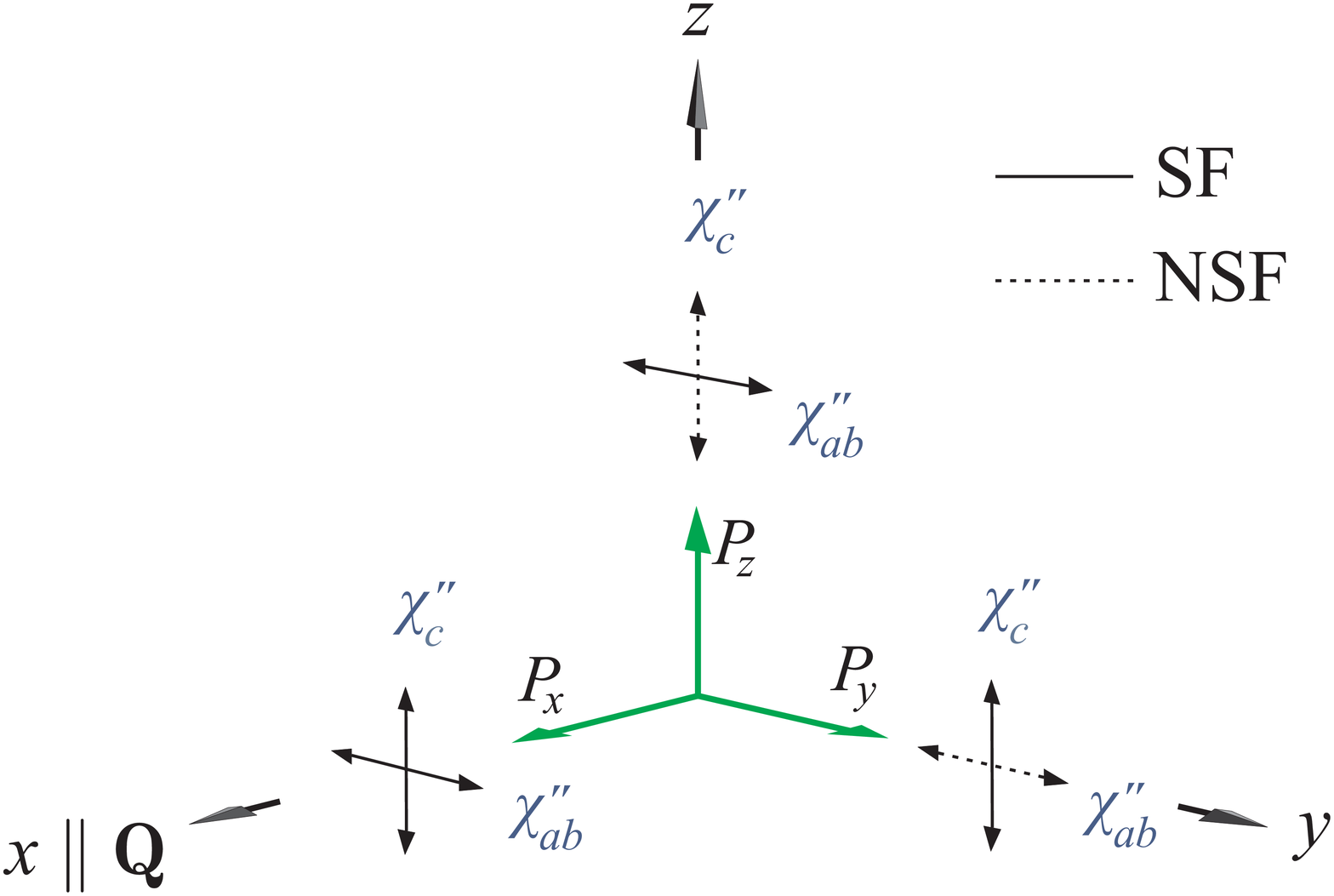}
\caption{(Color online) Diagram representing the axis convention used in this work. The scattering vector $\bf Q$ defines the $x$-axis, which in our experiment is always parallel to the $ab$ plane. Only the magnetic fluctuations perpendicular to $\bf Q$ are observed. The incident polarization vector ${\bf P}_{\rm i} = (P_x, P_y, P_z)$ is spin-flipped by the magnetic fluctuation component perpendicular to both $\bf Q$ and ${\bf P}_{\rm i}$. Components of the magnetic fluctuations which are perpendicular to $\bf Q$ but not ${\bf P}_{\rm i}$ appear in the non-spin-flip channel. Hence, a separation of the in-plane and out-of-plane susceptibilities $ \chi''_{ab}({\bf Q}, \omega)$ and $ \chi''_{c}({\bf Q}, \omega)$ can be achieved.
\label{fig:pol_ana}}
\end{figure}

In total, six neutron cross-sections were measured, denoted by $\sigma(x,\pm x)$, $\sigma(y,\pm y)$ and $\sigma(z,\pm z)$. The coordinate $x$ is taken along the scattering vector $\bf Q$, $z$ is perpendicular to the scattering plane (here $z \parallel c$) and $y$ completes the right-handed Cartesian system --- Fig.~\ref{fig:pol_ana}. The two indices in $\sigma$ refer to the direction of the neutron polarization before and after the sample, respectively.


\begin{figure}
\centering
\includegraphics[width=0.7\columnwidth,bb=-9 16  556 410]
{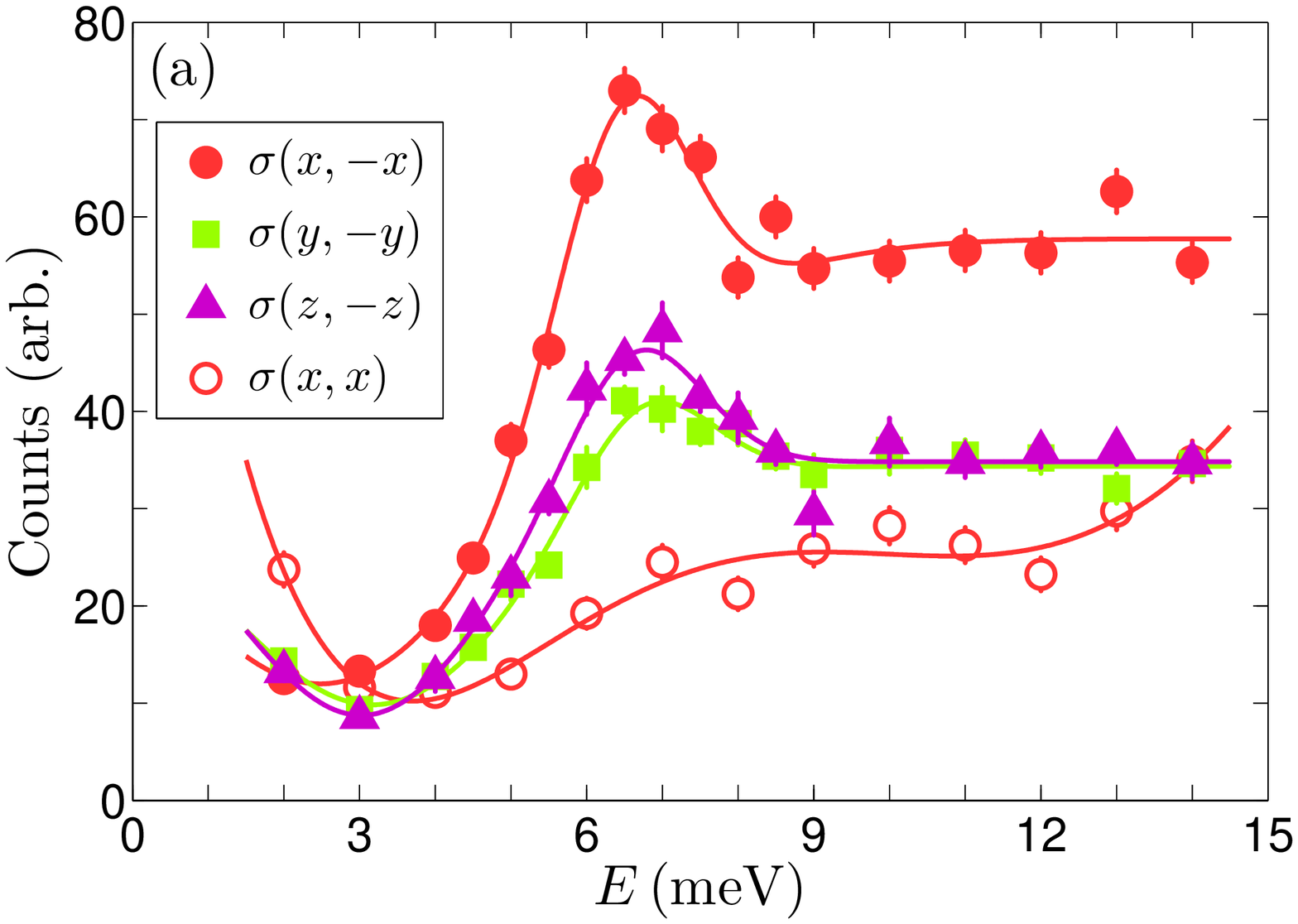}
\includegraphics[width=0.7\columnwidth,bb= 0 150 565 545]
{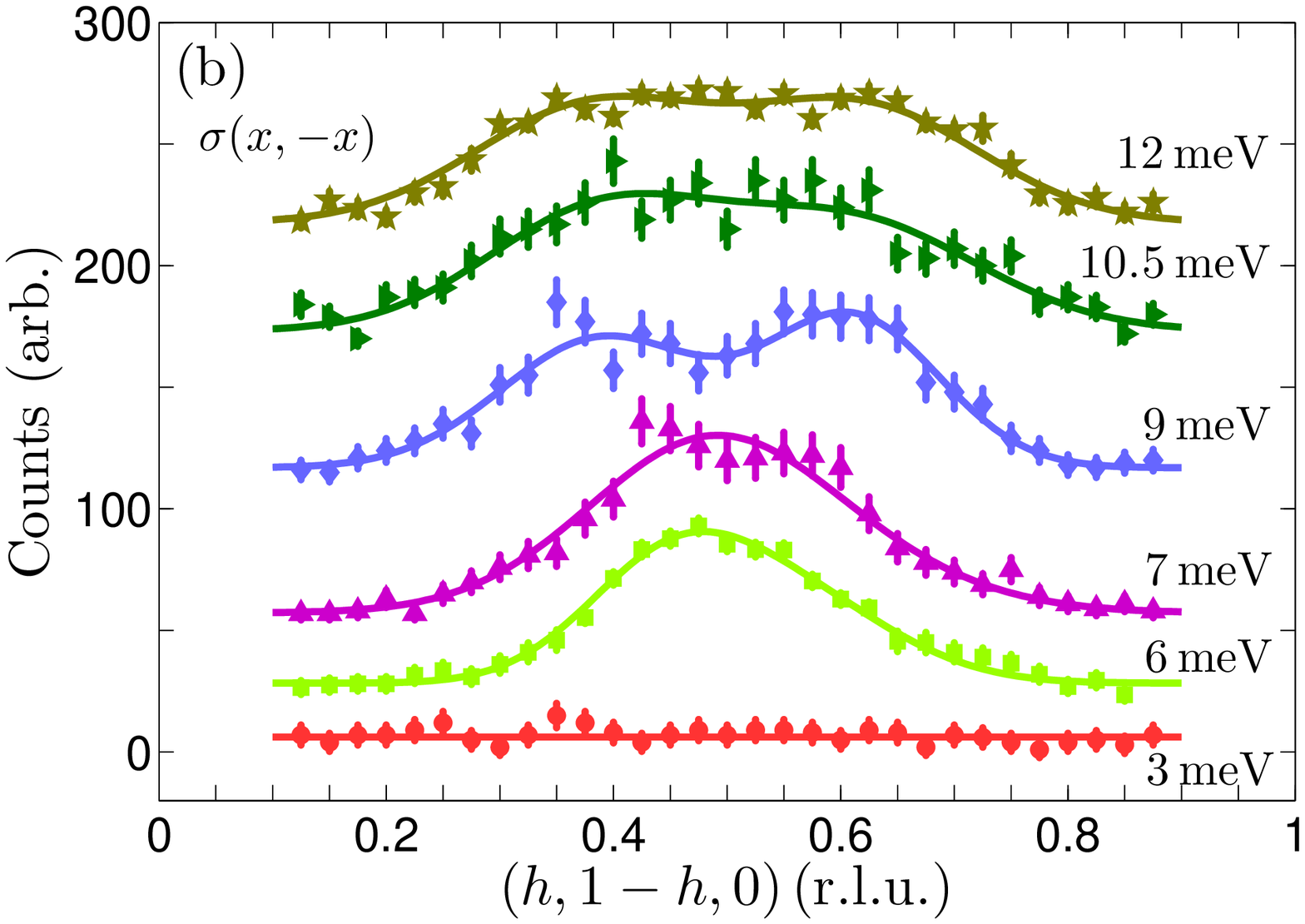}
\caption{(Color online) (a) Energy scans at ${\bf Q}_0 = (0.5,0.5,0)$ showing the SF channels which contain the magnetic scattering and the $\sigma(x,x)$ NSF channel which contains non-magnetic scattering. Lines are visual guides. (b) Wavevector scans along $(h,1-h,0)$ at energies of 3 to 12\,meV (displaced vertically) showing the $\sigma(x,-x)$ SF scattering. Solid lines show least-squares fits to the spectra assuming a Gaussian lineshape. Data in both (a) and (b) were recorded at a temperature of 2\,K.
\label{fig:My_Mz}}
\end{figure}

The crystal structure of FeSe$_{0.5}$Te$_{0.5}$ is tetragonal and so in general $ \chi''_{ab}({\bf Q}, \omega)$ can be different from $ \chi''_{c}({\bf Q}, \omega)$.  Longitudinal polarization analysis allows a complete separation of $ \chi''_{ab}({\bf Q}, \omega)$ and $ \chi''_{c}({\bf Q}, \omega)$ because of two properties of the magnetic scattering cross-section: (1) neutrons only scatter from spin fluctuations perpendicular to $\bf Q$, and (2) spin fluctuations perpendicular to the incident neutron polarization ${\bf P}_{\rm i}$ scatter in the spin-flip (SF) channel, while spin fluctuations parallel ${\bf P}_{\rm i}$ scatter in the non-spin-flip (NSF) channel.  With the geometry chosen for the present measurements the SF cross-sections are given by \cite{squires-book}
\begin{eqnarray}
  \sigma(x,-x) & \propto &  \chi''_{ab} + \chi''_{c} + B^{\rm SF} \nonumber\\
  \sigma(y,-y) &\propto&  \chi''_{c}         + B^{\rm SF} \nonumber\\
  \sigma(z,-z) &\propto&  \chi''_{ab}      + B^{\rm SF},
  \label{eq:SF}
\end{eqnarray}
and the NSF cross-sections
\begin{eqnarray}
  \sigma(x,x) &\propto& N + B^{\rm NSF}            \nonumber\\
  \hphantom{-}
  \sigma(y,y) &\propto&  \chi''_{ab}  + N + B^{\rm NSF} \hphantom{+\chi''_{c}}\nonumber\\
  \sigma(z,z) &\propto&  \chi''_{c}  + N + B^{\rm NSF},
  \label{eq:NSF}
\end{eqnarray}
where $N$ refers to the coherent nuclear cross-section, and $B^{\rm SF}$ and $B^{\rm NSF}$ are the SF and NSF backgrounds. To simplify the notation we omit the explicit dependence on $\bf Q$ and $\omega$ from now on. These scattering processes are represented in Fig.~\ref{fig:pol_ana}. The background was found to be independent of the polarization in the SF cross-sections to within experimental error from measurements at ${\bf Q} \approx (0.1,0.9,0)$ and $E\approx 6$\,meV.

Figure~\ref{fig:My_Mz}(a) shows energy scans performed at ${\bf Q}_0 =(0.5,0.5,0)$ in the three SF channels and in the $\sigma(x,x)$ NSF channel. The intensity in the $\sigma(x,x)$ channel is significant, highlighting the importance of using polarized neutron scattering to separate the nuclear contribution from the magnetic signal. From Eq.~\ref{eq:SF}, the $\sigma(x,-x)$ cross-section contains the total magnetic scattering. The scattering in this channel contains a peak at $\hbar\omega_0 \approx6.5$\,meV, corresponding to the spin resonance previously reported by unpolarized inelastic neutron scattering measurements in compounds of similar composition \cite{qiu-prl-2009,babkevich-jpcm-2010,argyriou-prb-2010,mook-prl-2010,
wen-prb-2010}. Figure~\ref{fig:My_Mz}(b) shows the $\sigma(x,-x)$ cross-section in wavevector scans along $(h,1-h,0)$ at selected energies. At 3\,meV only a flat background is evident. Above the resonance energy, steeply-rising incommensurate magnetic excitations are observed. Our results are consistent with unpolarized neutron scattering measurements on FeSe$_{0.5}$Te$_{0.5}$, Refs.~\onlinecite{argyriou-prb-2010,mook-prl-2010}.

\begin{figure}
\centering
\includegraphics[width=0.7\columnwidth]
{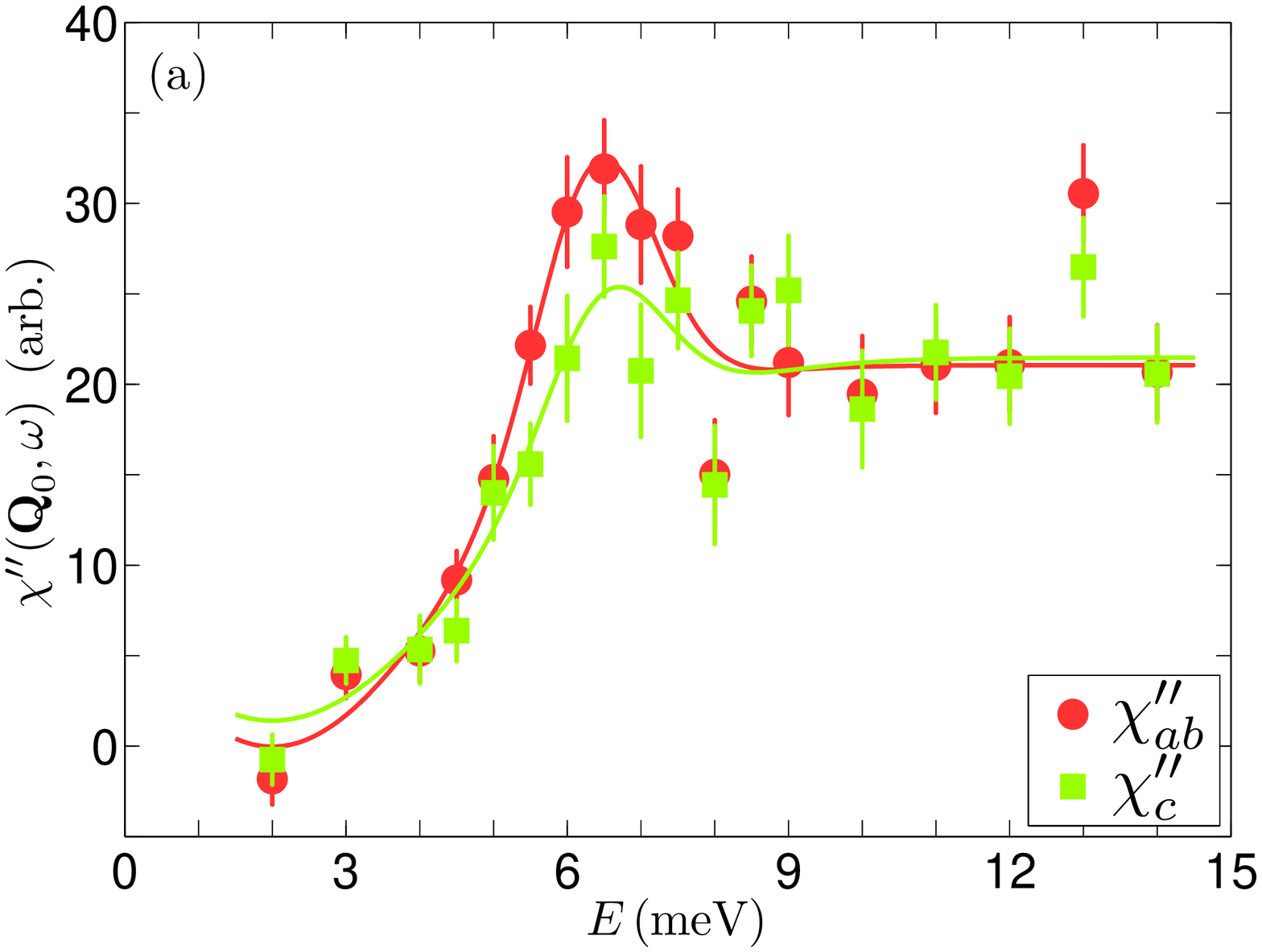}

\vspace{0.2cm}

\includegraphics[width=\columnwidth]
{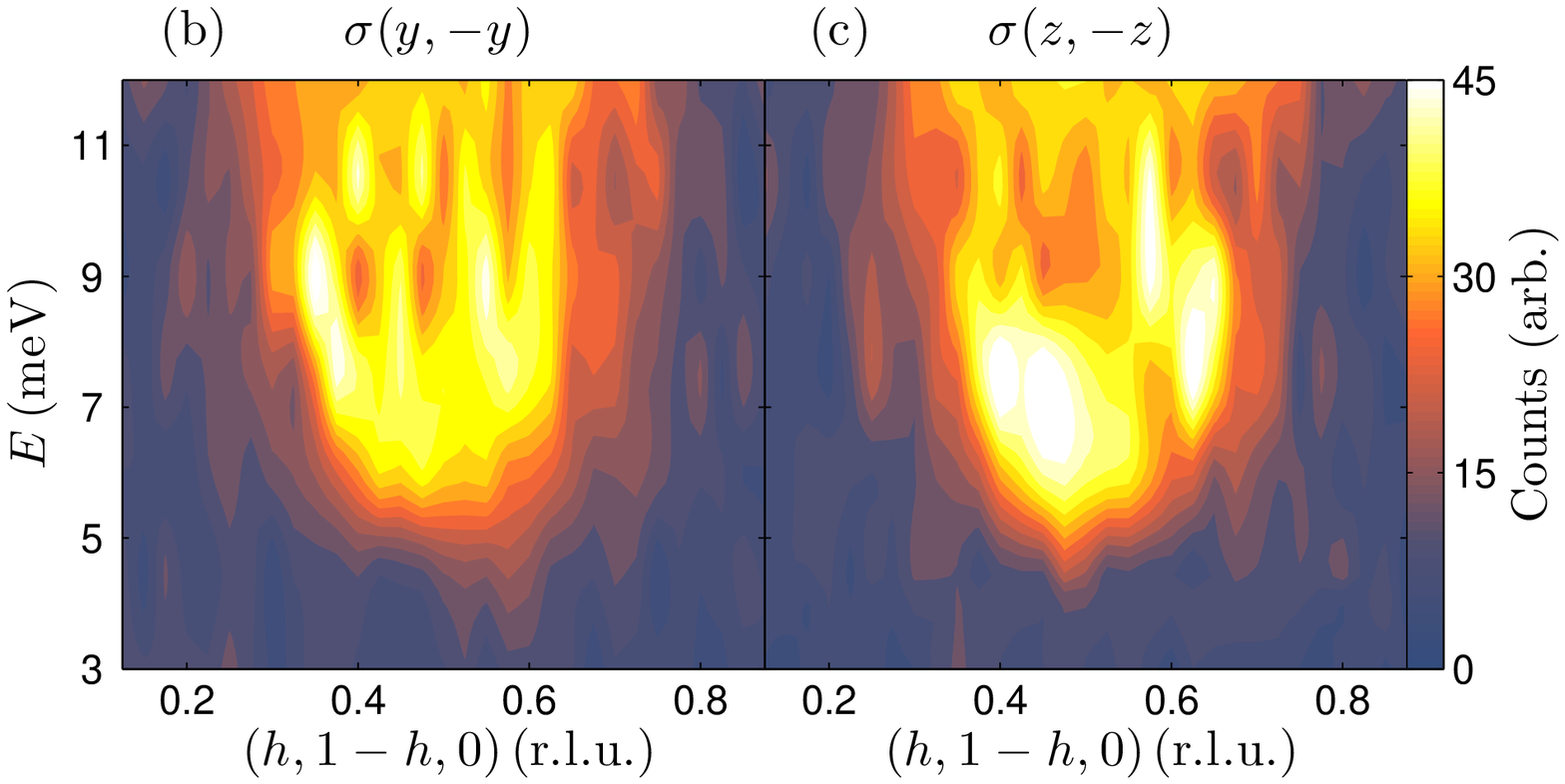}
\caption{(Color online) (a) Comparison of the scattering from in-plane ($\chi''_{ab}$) and out-of-plane ($\chi''_{c}$) magnetic fluctuations in FeSe$_{0.5}$Te$_{0.5}$. Solid lines through the data points are guides to the eye. (b) and (c) Intensity maps showing the cross-sections $\sigma(y,-y)$ and $\sigma(z,-z)$, which contain $ \chi''_{c}$ and $ \chi''_{ab}$, respectively. All the data in this figure were recorded at $T = 2$\,K.}
\label{fig:maps}
\end{figure}

The $\sigma(y,-y)$ and $\sigma(z,-z)$ SF channels, shown in Fig.~\ref{fig:My_Mz}(a), contain the magnetic scattering from out-of-plane and in-plane fluctuations, respectively --- see Eq.~\ref{eq:SF}. The signal in these channels is very similar throughout the energy range measured, both channels having a peak at the resonance energy. A small but statistically significant difference is observed between $\sigma(y,-y)$ and $\sigma(z,-z)$ on the resonance peak itself.  Using Eq.~\ref{eq:SF} we can eliminate the background contribution and separate the in-plane and out-of-plane components of magnetic scattering:  $ \chi''_{ab} \propto \sigma(x,-x) - \sigma(y,-y)$ and  $ \chi''_{c} \propto \sigma(x,-x) - \sigma(z,-z)$. Figure~\ref{fig:maps}(a) shows the result of this procedure. The resonance peak appears at the same energy to within an experimental error of 1\,meV in both $ \chi''_{ab}$ and $ \chi''_{c}$. The peak is slightly larger in $ \chi''_{ab}$.  Either side of the spin resonance energy the intensity is approximately the same for both channels.

The similarity between the $ \chi''_{ab}$ and $ \chi''_{c}$ components is emphasized in the color maps shown in Fig.~\ref{fig:maps}(b) and (c), which show the intensity distribution as a function of energy and wavevector along $(h,1-h,0)$. The data plotted in these maps are the $\sigma(y,-y)$ and $\sigma(z,-z)$ cross-sections, which contain the $ \chi''_{c}$ and $ \chi''_{ab}$ fluctuations, respectively.  The overall conclusion from all the $T=2$\,K data is that the low-energy spin fluctuations in FeSe$_{0.5}$Te$_{0.5}$ are isotropic ($\chi''_{ab} \approx  \chi''_{c}$) to within experimental error, except on the resonance peak itself where $ \chi''_{ab}$ is approximately 20\% larger than $ \chi''_{c}$.

\begin{figure}
\centering
\includegraphics[width=0.7\columnwidth]
{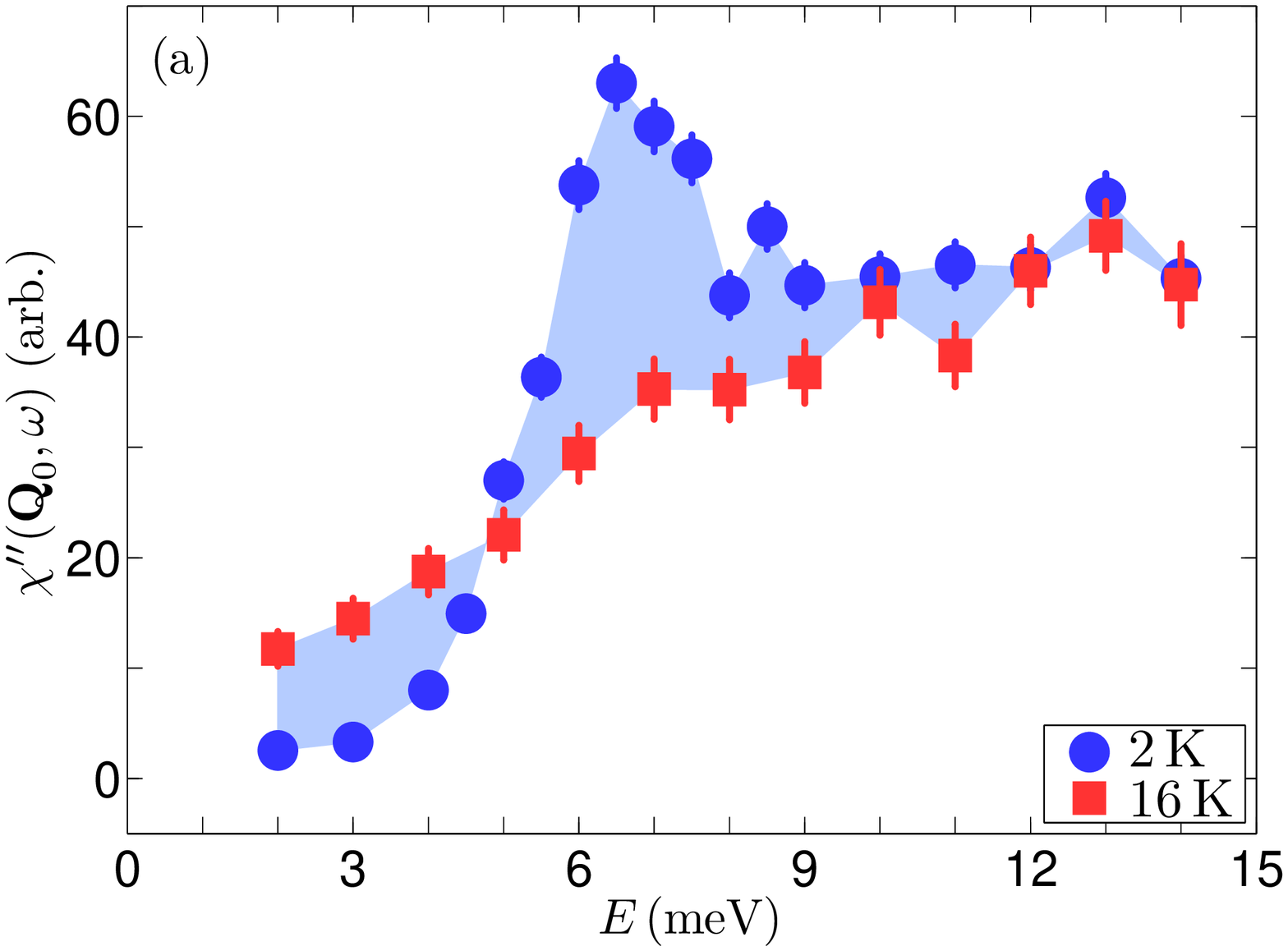}
\includegraphics[width=0.9\columnwidth]
{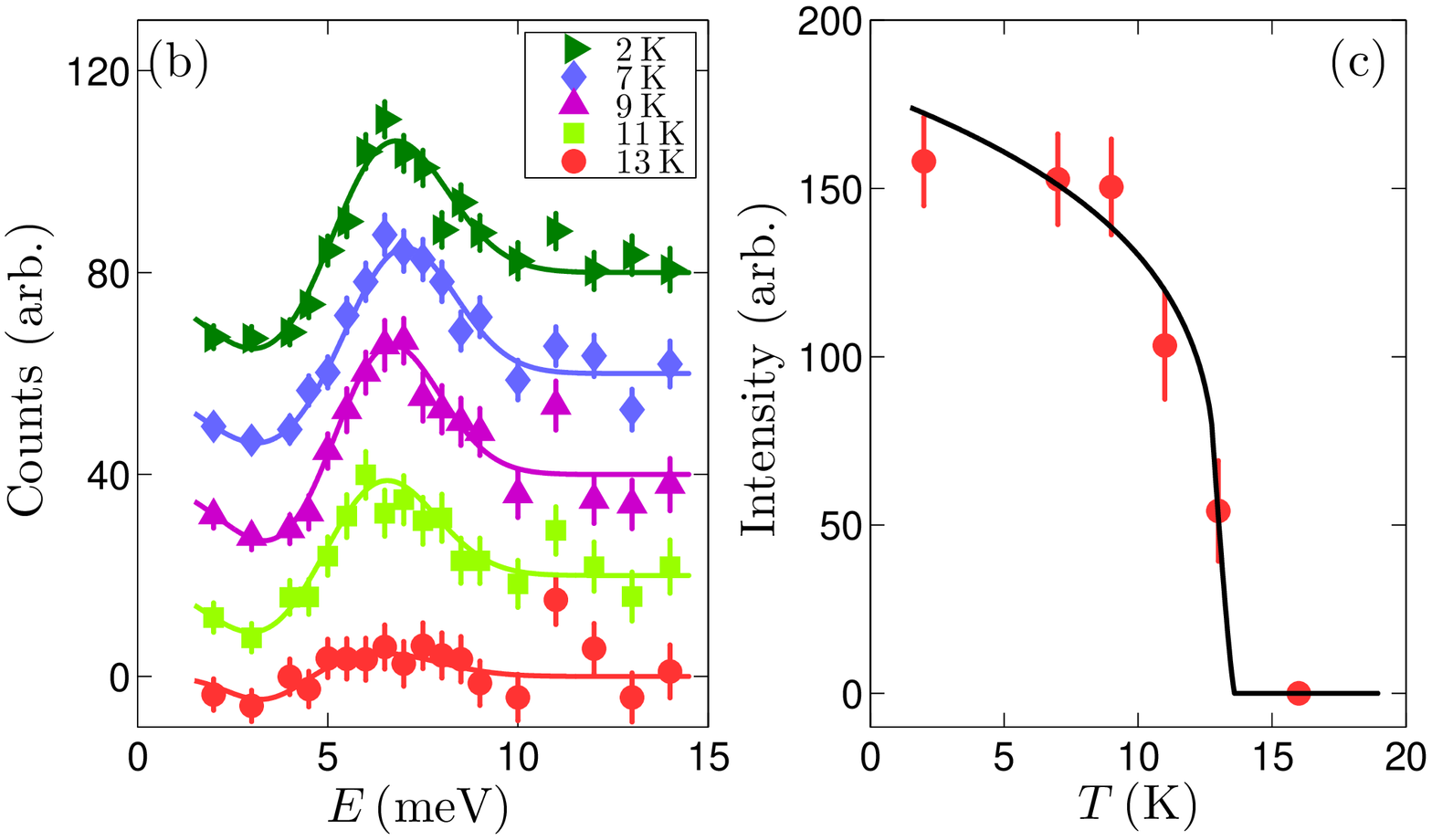}
\caption{(Color online) Temperature dependence of the spin resonance in FeSe$_{0.5}$Te$_{0.5}$. (a) Energy scans at ${\bf Q}_0 = (0.5,0.5,0)$ showing $\chi'' = \chi''_{ab}+\chi''_{c}$ at 2\,K and 16\,K. The shaded region highlights the change in the spectrum with temperature. (b) Energy scans at ${\bf Q}_0$ for a series of temperatures from 2 to 13\,K. The scans are displaced vertically. (c) Integrated intensity of the resonance peak as a function of temperature. Data recorded at 16\,K have been subtracted in panels (b) and (c). Solid lines are guides to the eye.
\label{fig:tdept}}
\end{figure}

Figure~\ref{fig:tdept} presents the results of measurements of the temperature dependence of the magnetic fluctuations at ${\bf Q}_0 = (0.5,0.5,0)$ in FeSe$_{0.5}$Te$_{0.5}$. Here we show data obtained from the $\sigma(x,-x)$ cross-section, which from Eq.~(\ref{eq:SF}) is proportional to $\chi''_{ab}({\bf Q}_0, \omega)+\chi''_{c}({\bf Q}_0, \omega)$. Because the measured intensity is proportional to $ \chi''({\bf Q},\omega)/[1 - \exp(-\hbar\omega/k_{\rm B}T)]$ we have multiplied the intensity by $1 - \exp(-\hbar\omega/k_{\rm B}T)$ to compare susceptibilities at different temperatures. We see that the resonance peak disappears above $T_{\rm c} = 14$\,K, while at higher energies the susceptibility remains essentially unchanged. At 16\,K we also observe an increased response below the spin gap. Figure~\ref{fig:tdept}(b) shows the temperature evolution of the $\sigma(x,-x)$ cross-section for temperatures from 2 to 13\,K. A scan measured at 16\,K (above $T_{\rm c}$) was subtracted to isolate the spin resonance contribution. Upon warming, the intensity of the spin resonance shows little change up to 9\,K. When the temperature approaches $T_{\rm c}$, the spectral weight diminishes and the spin-gap is gradually filled. Another notable feature is that the spin resonance does not shift to lower energies with increasing temperature, as one might expect if the spin resonance were simply caused by a gap which closes at $T_{\rm c}$ with temperature. From our measurements we conclude that the position and the energy width of the spin resonance are temperature independent up to at least $\sim0.8T_{\rm c}$. The lack of softening of the resonance energy with increasing temperature has also been found in FeSe$_{0.4}$Te$_{0.6}$ \cite{qiu-prl-2009}. Figure~\ref{fig:tdept}(c) shows the evolution of the integrated intensity of the spin resonance which behaves as an order parameter of the superconducting phase. In the vicinity of $T_{\rm c}$ measurements with higher precision are needed to obtain a more quantitative estimate of the renormalization of the inelastic intensity than is available from the present experiment.

The polarized-neutron data presented here go beyond what has hitherto been possible with unpolarized neutron scattering, and provide new insights into the magnetic excitations of FeSe$_{0.5}$Te$_{0.5}$. A superconducting wavefunction with purely $s^{\pm}$ pairing state would result in an isotropic spin resonance peak \cite{maier-prb-2008}. Our data suggests a small anisotropy, in the sense $ \chi''_{ab} > \chi''_{c}$. This small anisotropy cannot readily be explained by the usual anisotropic terms in the spin Hamiltonian since the magnetic scattering is isotropic above and below the resonance peak. It is possible, therefore, that the superconducting pairing function contains a minority component with a different symmetry. For example, a spin-triplet with sign-reversed $p$-wave gap is predicted to give a resonance in $ \chi''_{ab}$, but not in $ \chi''_{c}$ \cite{maier-prb-2008}.

The relatively small anisotropy in the spin resonance of FeSe$_{0.5}$Te$_{0.5}$, consistent with a dominant singlet--triplet excitation, is in stark contrast to the results of a study on BaFe$_{1.9}$Ni$_{0.1}$As$_2$, which revealed a highly anisotropic spin resonance with only the $ \chi''_{ab}$ component non-zero \cite{lipscombe-prb-2010}. The results also differ from the spin-ladder system Sr$_{14}$Cu$_{24}$O$_{41}$, which also has a resonance-like coherent singlet--triplet excitation \cite{lorenzo-prl-2010}. Firstly, the anisotropy is in the opposite sense (in Sr$_{14}$Cu$_{24}$O$_{41}$ the out-of-plane fluctuations are stronger than the in-plane fluctuations), and second, the anisotropy in Sr$_{14}$Cu$_{24}$O$_{41}$ is observed over a range of energies not just on the peak \cite{boullier-physicab-2004}.

Recently, polarized-neutron scattering measurements have been performed on YBa$_2$Cu$_3$O$_{6+x}$ \cite{headings-arxiv-2011,regnault-unpub}. The spin resonance in YBa$_2$Cu$_3$O$_{6.9}$ at 40\,meV, corresponding to the odd-parity mode, was found to be quasi-isotropic to within the precision of the measurements. This implies that the resonance peak is predominantly a singlet--triplet excitation in both FeSe$_{0.5}$Te$_{0.5}$ and YBa$_2$Cu$_3$O$_{6+x}$. Furthermore, the resonance peaks in these materials do not soften appreciably as the temperature is increased towards $T_{\rm c}$ (see Fig.~\ref{fig:tdept} and Ref.~\onlinecite{fong-prb-1996}).  These similarities suggest that the superconducting states in the cuprates and Fe-based superconductors have some general features in common.

\bibliographystyle{apsrev4-1}

\bibliography{biblio}

\end{document}